# Modeling capacitor voltage in an RC circuit from data: Construction, breakdown, and revision of an empirical model

Kazumasa Kushida*

Department of Educational Collaboration, Osaka Kyoiku University, Kashiwara, Osaka 582-8582, Japan

**Abstract**

We describe a guided introductory laboratory activity in which students construct empirical models of capacitor voltage in RC circuits from measured input–output data before circuit theory is introduced. Students may use what they know about resistors, Ohm's law, and conservation of charge and energy, but prior knowledge of capacitor behavior is not used as a reason for adopting a model. The circuits use a 1.5-V battery, resistors, and electric double-layer capacitors with nominal capacitances of 4.7 and 10 F. Their slow responses are recorded with handheld voltmeters and smartphone video, and the capacitance values are presented to students with the unit withheld. To compare circuits, students rescale time by the product of the nominal resistance and capacitance, as proposed by the instructor. On this coordinate, plots of the logarithm of the normalized discharging voltage are approximately straight lines through the origin, providing empirical models. For charging, the corresponding logarithmic plots are clearly curved, and the discharging model breaks down. Using the remaining distance from the final value as the dependent variable gives approximately straight logarithmic plots and provides a basis for revising the model. Rewriting the discharging model in equivalent differential and integral forms allows students to compare changes in charge-like and energy-like quantities with the charge transferred through the resistor and the energy dissipated in it. Standard circuit theory, introduced afterward, describes charging and discharging in a common form and gives these quantities their conventional meanings.

* *Corresponding Author*: Kazumasa Kushida
* *E-mail*: kkushi@cc.osaka-kyoiku.ac.jp

## I. INTRODUCTION

Building models is a central part of physics.[1] In one sequence of introductory laboratory activities, established relations were rewritten in linear form so that a physical parameter could be obtained from a fitted straight line.[2] Another approach asked students to plan experiments and construct relations from their own data before the corresponding theory was introduced.[3] In an upper-division laboratory activity, model construction, prediction, comparison with data, and refinement were made explicit learning goals.[4] Deviations of RC decay data from a single exponential have also been used in an advanced laboratory to engage students in data validation and model construction.[5] Such model-building efforts accord with the AAPT laboratory recommendations, which emphasize constructing knowledge from real data without relying on an outside authority and understanding the limitations of models.[6]

The present work reverses the order of the conventional RC experiment. The capacitor in an RC circuit is treated as a black box: its voltage response is characterized through observable input–output relations, and standard series-RC theory is not used to prescribe the model. For discharging, rescaling time provides a common coordinate for comparing different circuits. Graphical linearization with a logarithmic plot, as in Ref. 2, provides a basis for constructing an empirical model. For charging, the corresponding logarithmic plots are clearly curved, and the model breaks down. Using the remaining distance from the final value as the dependent variable gives approximately straight logarithmic plots and provides a basis for revising the model. This sequence makes the construction, breakdown, and revision of empirical models explicit. This activity is guided rather than discovery-based: the instructor proposes or instructs its key steps. Physical meanings are examined in the post-experimental discussion, where standard series-RC theory is introduced.

The activity is intended for an introductory undergraduate laboratory in which students learn how a capacitor behaves in an RC circuit. Its goal is not to verify the standard exponential formulas, but to lead students through the process of constructing an empirical model from their data, observing its breakdown, and revising it.

## II. EXPERIMENTAL SETUP AND INSTRUCTIONAL DESIGN

### A. Apparatus and measurements

The activity is conducted over two 90-min sessions in an introductory undergraduate physics laboratory with approximately 30 students working in groups of two or three. Students assemble series RC circuits on a solderless breadboard by consulting the circuit diagram and comparing their work with a circuit assembled by the instructor. Each circuit comprises a 1.5-V zinc–carbon dry-cell battery, a resistor (R = 4.7, 10, or 20 Ω; tolerance ±5% for 4.7 and 10 Ω and ±1% for 20 Ω), and an electric double-layer capacitor (EDLC; C = 4.7 or 10 F; tolerance ±20%). The capacitor voltage $V$ is measured with a digital voltmeter (YF-1000, TENMARS ELECTRONICS CO., LTD., Taipei, Taiwan) with a resolution of 1 mV. The voltmeter display is filmed with a smartphone, and readings are transcribed from the video.

The EDLCs are not modeled in detail; they are chosen because their large capacitances yield response times of minutes. The objective is not precision measurement but the identification of trends that can be compared across circuits and used to construct an empirical model. Charging and discharging are measured consecutively in a single run. For charging, $E$ denotes the last reading of the charging record; for discharging, it denotes the initial voltage of the discharging record.

Five combinations of nominal values are used: (R, C) = (4.7 Ω, 4.7 F), (4.7 Ω, 10 F), (10 Ω, 4.7 F), (10 Ω, 10 F), and (20 Ω, 10 F). Each group measures charging and discharging for two of the first three combinations and obtains data for the third from another group; the instructor supplies data for the remaining two. Figures 1 and 2 show representative measurements obtained under the same five conditions; they are not pooled student measurements.

**B. Instructional guidance**

The instructor provides the circuit diagram and the measurement procedure but withholds the circuit differential equation, its exponential solutions, and the interpretation of $RC$ as a time constant. Students may freely use what they know about voltage, current, resistance, batteries, Ohm's law, and conservation of charge and energy. Students may already know the properties of capacitors, the circuit differential equation, and its exponential solutions, but this knowledge is not used as a reason for adopting an empirical model; the model is adopted because it describes the measured input-output relation over a stated range.

Before the main measurements, students observe qualitatively in a preliminary experiment that the voltage changes more slowly for larger capacitances and resistances. Because the observations concern how the capacitor voltage changes with time, the elapsed time $t$ is treated as the independent variable and the voltage $V(t)$ across the capacitor as the dependent variable. By analogy with a rechargeable battery, students call the two procedures charging and discharging. From the main measurements on, the capacitances are presented as $\Gamma = 4.7\ [X]$ and $10\ [X]$ in place of $C = 4.7\ \mathrm{F}$ and $10\ \mathrm{F}$. The symbol $\Gamma$ does not denote an additional physical quantity; $\Gamma$ carries the same numerical values, and $[X]$ marks a unit that is not named, so that $\Gamma$ is treated as a dimensional quantity whose unit is to be inferred. The instructor then asks, "From the preliminary experiment, how is the magnitude of $\Gamma$ related to the rate at which the voltage changes?"

During the analysis, the instructor proposes or instructs each key step, such as the normalization of the voltage, the scaling of time, the graphical linearization, and the change of the dependent variable for charging. Questions are posed before and after each step so that students connect it with their own data and retain it as a method. When a group cannot carry out a step, the instructor shows an example, including the instructor's data file for Figs. 1 and 2, and the group follows it. A functional form fitted to an observed relation over a stated range of the data is treated as an empirical model.

## III. CONSTRUCTING AN EMPIRICAL MODEL FOR DISCHARGING

### A. Comparing circuits and choosing a time coordinate

The discharging voltage of each run is normalized by its initial voltage, $v = V/E$, with $E =$

$V(0)$, so that all curves start from $v = 1$ [Fig. 1(a)]. $E$ differs by at most 2 mV from the voltage at the end of the preceding charging (Sec. II A). The curves fall at clearly different rates: the half-decay time, at which $v = 0.5$, ranges from approximately 12 s for $(4.7\ \Omega,\ 4.7\ [X])$ to 136 s for $(20\ \Omega,\ 10\ [X])$.

The decay becomes slower when $R$ or $\Gamma$ is increased. The instructor therefore proposes scaling time by a combined quantity that increases with either $R$ or $\Gamma$, so that the voltage responses of different circuits can be compared. Both $R + \Gamma$ and $R\Gamma$ satisfy this condition, but a sum requires its terms to have the same dimension, and there is no basis for judging whether $\Gamma$, expressed in $[X]$, has the same dimension as $R$. The product $R\Gamma$ is therefore tried first. The aim of this proposal is not for students to independently invent an unfamiliar scaling transformation, but for them to apply one, evaluate what it reveals in measured data, and acquire it as a method that they may draw upon in future modeling tasks.

Students then replot the data against $T = t/(R\Gamma)$, computed from the nominal values [Fig. 1(b)]. On this coordinate, the five curves lie close to one another, although they do not coincide completely. $T$ therefore serves as a common coordinate on which the data sets are compared. At this stage, $R\Gamma$ is used only to construct this coordinate, and no physical meaning has yet been assigned to it. An alternative scaling by a time obtained from each data set is described in the Appendix.

**B. A straight-line model**

Asked what function the curves in Fig. 1(b) resemble, students usually answer that they look exponential. To test this, the instructor proposes graphical linearization: if the data follow an exponential function, a plot of $\ln v$ versus $T$ should be a straight line [Fig. 1(c)]. The instructor also points out that the argument of a logarithm must be dimensionless, as ensured by the normalization $v = V/E$. In Fig. 1(c), each data set follows an approximately straight line through the origin over $0 < T \leq 3$, with slight deviations. Students therefore adopt the empirical model

$$\ln v = -k_{\mathrm{d},i} T. \qquad (1)$$

Here, $k_{\mathrm{d},i}$ is obtained for data set $i$ by an origin-constrained least-squares fit over this range. The origin constraint follows from the definition $E = V(0)$, which gives $v = 1$, and hence $\ln v = 0$, at $T = 0$. No additional parameters are introduced to account for the slight deviations.

Introducing $k_{\mathrm{d},i}$ requires a decision about the dimensions of $T$ and $k_{\mathrm{d},i}$. The only clue is that $\ln v$ is dimensionless, so that the product $k_{\mathrm{d},i} T$ must also be dimensionless. Either $T$ is dimensionless and $k_{\mathrm{d},i}$ is a pure number, or $T$ carries a dimension and $k_{\mathrm{d},i}$ the inverse one. The instructor proposes the simpler choice, in which both $T$ and $k_{\mathrm{d},i}$ are dimensionless. A dimensional quantity acquires meaning only through comparison with a relevant quantity of the same dimension.[7] Taking $T = t/(R\Gamma)$ to be dimensionless thus amounts to comparing $t$ with $R\Gamma$, so that $R\Gamma$ has the dimension of time. The unit of $\Gamma$ is examined in Sec. V A.

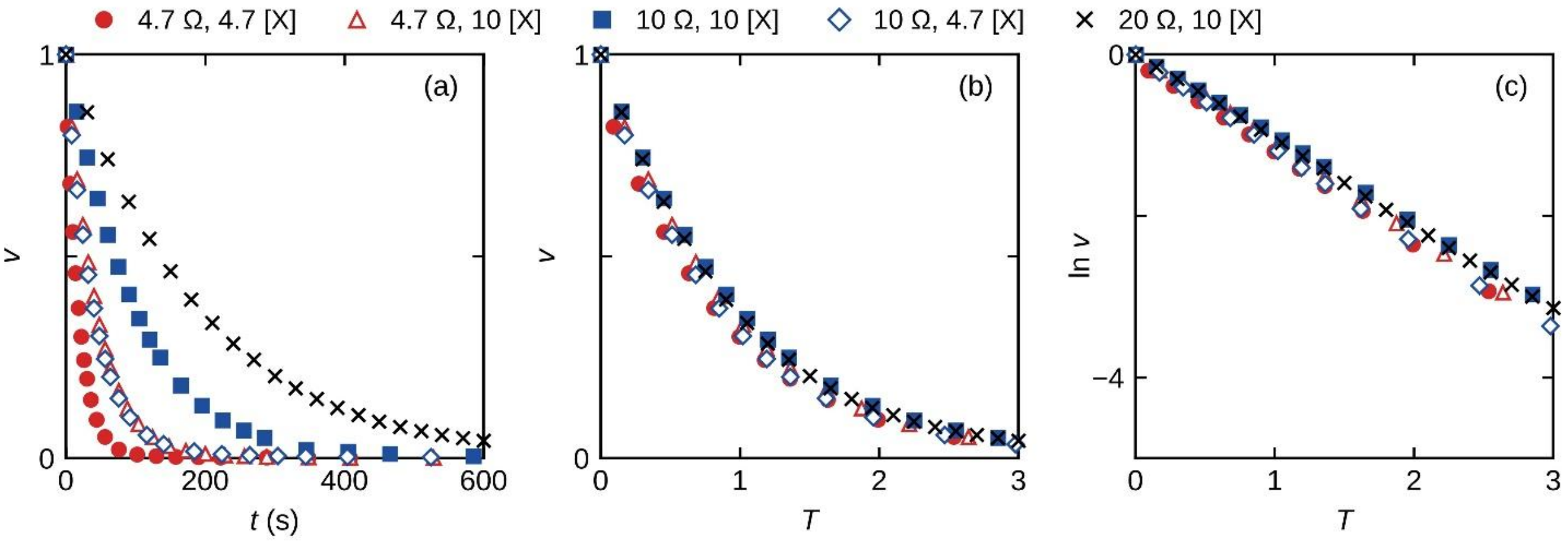

**FIG. 1.** Representative discharging data for the five combinations of $R$ and $\Gamma$ (Sec. II A). (a) Normalized voltage $v = V/E$ versus time $\Gamma$, with $E = V(0)$. (b) The same data versus $T = t/(R\Gamma)$, computed from the nominal values. (c) $\ln v$ versus $T$ over $0 < T \leq 3$. The legend gives $\Gamma$ with the unnamed unit [X] used in the activity.

## IV. CHARGING: BREAKDOWN AND REVISION

### A. Applying the discharging model

Students next examine charging in the same five circuits. The charging voltage of each run is normalized as $v = (V - V_0)/(E - V_0)$, where $V_0$ is the residual voltage at $t = 0$, at most 4 mV, and $E$ is the voltage at the end of charging (Sec. II A). The normalized voltage $v$ therefore increases from 0 toward 1 [Fig. 2(a)]. On the coordinate $T$ [Fig. 2(b)], the charging curves again lie close to one another, although they do not coincide completely.

Students then plot $\ln v$ versus $T$ for the charging data, as they did for discharging [Fig. 2(c)]. For every charging data set, $\ln v$ increases steeply at small $T$ and bends toward zero as $T$ increases. The straight-line model constructed for discharging therefore breaks down for charging. Because the time coordinate $T$ still reduces the spread among the curves, $T$ is retained, and the dependent variable $v$ is reconsidered.

### B. Revising the dependent variable

The instructor asks, “What values does $v$ approach in discharging and charging?” During discharging, $v$ decreases toward its final value, 0, so that $v$ itself is the remaining distance from the final value; Fig. 1(c) shows that its logarithm decreases approximately linearly with $T$. During charging, $v$ increases toward its final value, 1. By analogy with discharging, the instructor proposes taking the remaining distance from the final value, $1 - v$, as the dependent variable.

In a plot of $\ln(1 - v)$ versus $T$ [Fig. 2(d)], each data set follows an approximately straight line through the origin over $0 < T \leq 3$, with slight deviations. As in Sec. III B, students therefore adopt the revised empirical model with a dimensionless coefficient $k_{\mathrm{c},i}$,

$$\ln(1 - v) = -k_{\mathrm{c},i} T. \qquad (2)$$

Here, $k_{\mathrm{c},i}$ is obtained for data set $i$ by an origin-constrained least-squares fit over this range. The origin constraint follows from the charging normalization, which gives $v = 0$, and hence $\ln(1 - v) = 0$, at $T = 0$.

In both charging and discharging processes, the logarithm of the remaining distance from the final value decreases approximately linearly with $T$ over $0 < T \leq 3$. For discharging, this distance is $v$; for charging, it is $1 - v$.

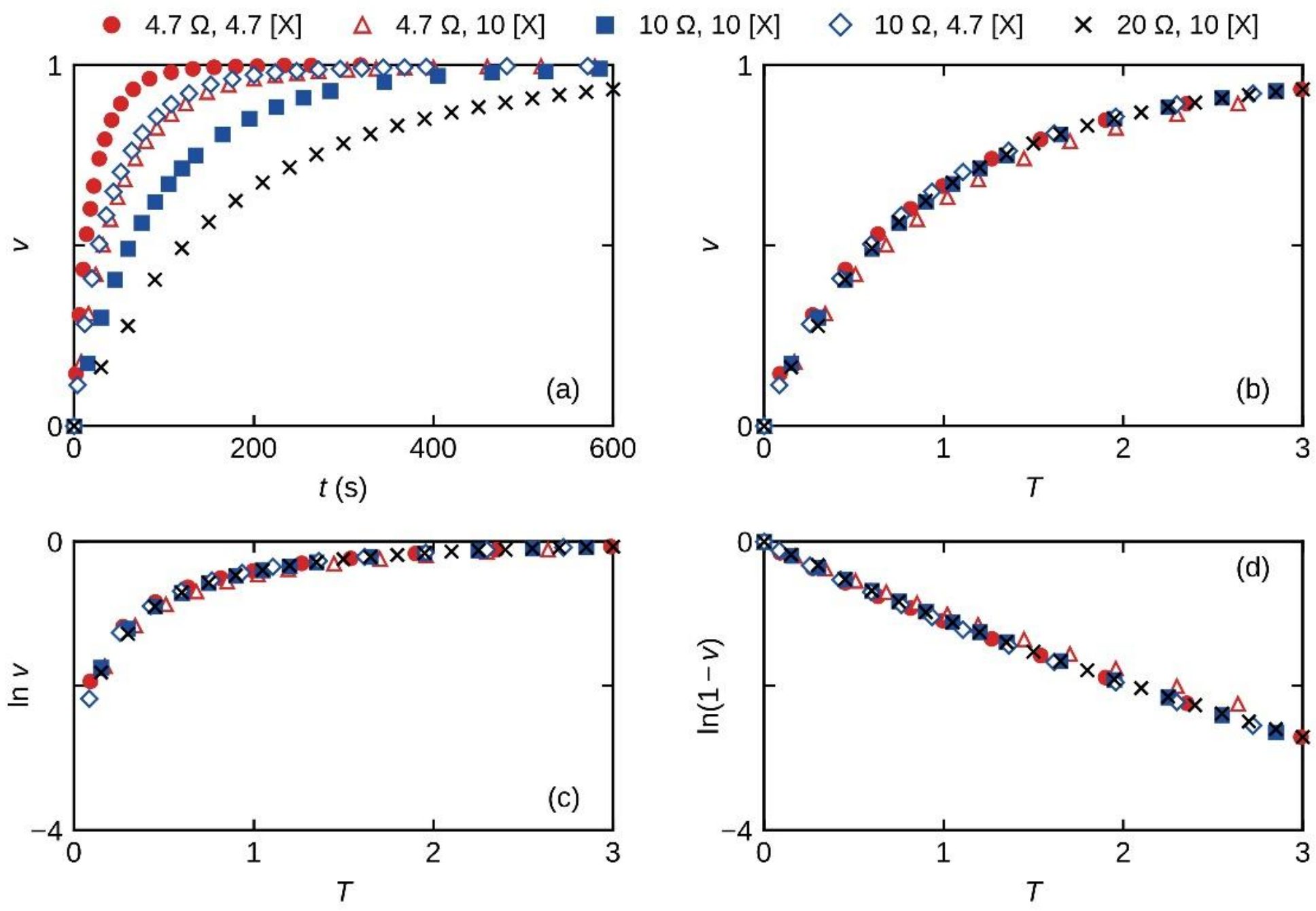

**FIG. 2.** Representative charging data for the same five combinations of $R$ and $\Gamma$ (Sec. II A). (a) Normalized voltage $v = (V - V_0)/(E - V_0)$ versus time $t$, where $V_0$ is the voltage at $t = 0$ and $E$ is the voltage at the end of charging. (b) The same data versus $T = t/(R\Gamma)$. (c) $\ln v$ versus $T$ over $0 < T \leq 3$; the data are clearly curved and do not follow straight lines. (d) $\ln(1 - v)$ versus $T$ over $0 < T \leq 3$. The legend is the same as in Fig. 1.

## V. POST-EXPERIMENTAL DISCUSSION

### A. Interpreting the empirical models before circuit theory

The post-experimental discussion is led by the instructor. When the results of the groups are compared, the discharging plots of $\ln v$ and the charging plots of $\ln(1-v)$ follow approximately straight lines for every group, but the coefficients $k_{\mathrm{d},i}$ and $k_{\mathrm{c},i}$ differ from one data set to another. The discussion does not assign physical significance to these differences; it turns instead to what the empirical model for discharging implies physically.

Students are guided to rewrite Eq. (1) in two equivalent forms. The following discussion applies to each data set separately, and the subscript $i$ is omitted. The rewriting uses only Eq. (1), the definitions $v = V/E$ and $T = t/(R\Gamma)$, and elementary calculus; no new measurement, fit, or circuit model is introduced. Differentiating Eq. (1) with respect to $t$ gives

$$-\frac{d(\Gamma V)}{dt} = \frac{k_d V}{R}. \qquad (3)$$

Multiplying Eq. (3) by $V$ and integrating from 0 to $t$ gives

$$\frac{\Gamma V^2(0)}{2} - \frac{\Gamma V^2(t)}{2} = k_d \int_0^t \frac{V^2(t')}{R} dt'. \qquad (4)$$

With the initial condition $v(0) = 1$, Eqs. (1), (3), and (4) are equivalent forms of the same empirical model. They are therefore subject to the same approximation and the same range, $0 < T \leq 3$, as Eq. (1).

The right-hand sides of Eqs. (3) and (4) contain quantities that students already know: the current $V/R$ through the resistor and the power $V^2/R$ dissipated in it. The left-hand sides contain $\Gamma V$ and $\Gamma V^2/2$, whose meanings have not yet been established. Because $T = t/(R\Gamma)$ is taken to be dimensionless (Sec. III B), $R\Gamma$ has the dimension of time; since $R$ is measured in ohms, [X] must be s/Ω. It follows that $\Gamma V$ has the unit A·s, that of charge, and that $\Gamma V^2/2$ has the unit of energy. The quantities $\Gamma V$ and $\Gamma V^2/2$ are therefore regarded as charge-like and energy-like quantities, respectively, both associated with the capacitor.

The internal mechanism of the capacitor remains unspecified, and in the discharging circuit the capacitor is connected only to the resistor. Equations (3) and (4) thus compare changes in the two inferred quantities with the charge transferred through the resistor and the energy dissipated in it.[7] Conservation of charge and energy is then brought into the discussion as prior physical knowledge, not as a conclusion from the fitted values. If $\Gamma V$ and $\Gamma V^2/2$ are identified with the charge and energy of the capacitor, conservation requires the rate of decrease of $\Gamma V$ to equal the current through the resistor and the decrease in $\Gamma V^2/2$ to equal the energy dissipated there. If $k_d > 1$, $\Gamma V$ would decrease faster than charge passes through the resistor, and the decrease in $\Gamma V^2/2$ would exceed the energy dissipated in it. If $k_d < 1$, charge would pass through the resistor faster than $\Gamma V$ decreases, and the energy dissipated in the resistor would exceed the decrease in $\Gamma V^2/2$.

For charging, the corresponding accounting is not pursued. The battery voltage under load was not measured and need not equal $E$, so $(E - V)/R$ cannot be identified with the charging current.

**B. Introducing circuit theory**

Standard series-RC theory is now introduced. $\Gamma$ is interpreted as the nominal value of the capacitance $C$, and $[X]$ is identified as the farad [F]. In the ideal circuit model, $RC$ is the time constant, $CV$ is the stored charge, and $CV^2/2$ is the stored energy. When the voltage is normalized by the initial and final values of the ideal model and time is scaled as $T_{\text{th}} = t/(RC)$, the charging and discharging solutions take the common form

$$\ln | v - v_{\infty} | = -T_{\text{th}}, \qquad (5)$$

where $v_{\infty} = 0$ for discharging and $v_{\infty} = 1$ for charging. Equation (5) has the functional forms of Eqs. (1) and (2) with a unit coefficient. Because $T$ is computed from nominal component values, whereas $T_{\text{th}}$ is defined using the parameters $R$ and $C$ of the ideal model, Eq. (5) does not imply that the fitted $k_d$ or $k_c$ equals unity. The theory assigns conventional meanings to the quantities identified in the empirical discussion. It is not used to evaluate or adjust the fitted coefficients; the quantitative relation between these coefficients and the ideal model lies outside the scope of the activity.

## VI. CONCLUSIONS AND IMPLICATIONS FOR INSTRUCTION

We have proposed an introductory laboratory activity in which the capacitor in a series RC circuit is treated as a black box and students construct an empirical model of its voltage response from their own data. Normalizing the voltage and scaling time by the product of the nominal values of resistance and capacitance allow students to compare the responses of different circuits on a common coordinate. Graphical linearization with a logarithmic plot then provides a basis for constructing the model.

The model constructed for discharging breaks down for charging. Its revision rests on an observation from discharging: the normalized voltage is itself the remaining distance from its final value, and the logarithm of this distance decreases approximately linearly with scaled time. Taking the corresponding distance, $1 - v$, as the dependent variable for charging gives an approximately linear logarithmic plot, suggesting that the revised functional form is appropriate over the analyzed range. This sequence of construction, breakdown, and revision follows, on a small scale, the pattern of epistemic iteration, in which each stage of knowledge is built on and used to correct the preceding stage.[8]

In this black-box approach, the physical interpretation of the capacitor response is not assigned in advance; it is suggested within the constraints of the mathematical form given to the input–output relation. In the present activity, a time scale and charge-like and energy-like quantities are identified in this way before circuit theory is introduced, and the theory then gives them their conventional meanings.

We have not measured student learning from the activity; that question requires a separate study.

## AUTHOR NOTE

## ACKNOWLEDGMENTS

This work was supported by JSPS KAKENHI Grant Number JP26K06064.

## AUTHOR DECLARATIONS

The author has no conflicts to disclose. The data that support the findings of this study are available from the author upon reasonable request.

## APPENDIX: SCALING BY A TIME OBTAINED FROM EACH DATA SET

An alternative to scaling by $R\Gamma$ is to scale time by a time $\tau_i$ obtained from each data set, for example, the time at which the tangent to the curve of $v$ versus $t$ at $t = 0$ intersects the time axis for discharging or the line $v = 1$ for charging. This approach uses only the measured response, but the relation of $\tau_i$ to $R$ and $\Gamma$ is then neither specified nor identified by the empirical model and would require a further empirical step. Scaling by $R\Gamma$ builds this relation into the common coordinate from the outset, and the inferences in Secs. III B and V A rest on it. The present activity therefore adopts scaling by $R\Gamma$.

Press, New York, 2004), pp. 224–228.